\documentclass[submission, Phys]{SciPost}

\usepackage[utf8]{inputenc} 
\usepackage[T1]{fontenc} 	
\usepackage[english]{babel} 

\usepackage[bitstream-charter]{mathdesign}
\usepackage{geometry} 		
\usepackage{amsmath} 		
\usepackage{mathtools} 		
\usepackage{float} 			
\usepackage{graphicx} 		
\usepackage{tabularx} 		
\usepackage{booktabs} 		
\usepackage{color, xcolor} 	
\usepackage{pdfpages} 		
\usepackage{extarrows} 		
\usepackage{multirow} 		
\usepackage{multicol} 		
\usepackage{enumitem} 		
\usepackage{xspace} 		
\usepackage{stackrel} 		
\usepackage{tikz} 			
\usepackage{braket} 		
\usepackage{bm} 			
\usepackage{tensor} 		
\usepackage{slashed} 		
\usepackage{siunitx} 		
\usepackage{lastpage} 		
\usepackage{cite} 			
\usepackage[normalem]{ulem} 
\usepackage{fontawesome} 	
\usepackage{tocloft} 		
\usepackage{titlesec} 		
\usepackage{doi} 			
\usepackage{hyperref} 		
\usepackage[most]{tcolorbox} 					
\usepackage[nameinlink, capitalize]{cleveref} 	
\usepackage[nottoc, notlot, notlof]{tocbibind} 	
\usepackage[ruled, vlined]{algorithm2e} 		
\usepackage{makecell}
\usepackage[makeroom]{cancel}
\usepackage{feynmf}

\makeatletter
\def\BState{\State\hskip-\ALG@thistlm}
\makeatother

\makeatletter
\@ifundefined{pdfoutput}{}{\DeclareGraphicsRule{*}{mps}{*}{}}
\makeatother

\makeatletter
\DeclareRobustCommand*{\bfseries}{%
   \not@math@alphabet\bfseries\mathbf
   \fontseries\bfdefault\selectfont
   \boldmath
}
\makeatother

\hypersetup{
	pdftitle={Agents of Discovery},
	pdfauthor={Diefenbacher et al.},
	colorlinks=true, 			
	linkcolor={red!50!black}, 	
	citecolor={blue!50!black}, 	
	urlcolor={blue!80!black} 	
} 

\DeclareSymbolFont{usualmathcal}{OMS}{cmsy}{m}{n}
\DeclareSymbolFontAlphabet{\mathcal}{usualmathcal}

\SetArgSty{textnormal}
\SetKwComment{Comment}{{\small\#}~}{}
\SetCommentSty{mycommfont}

\setitemize{itemsep=0pt, parsep=0pt} 				
\setenumerate{itemsep=0pt, parsep=0pt} 				
\setitemize{itemsep=2pt,topsep=2pt,parsep=0pt,partopsep=0pt,leftmargin=*}
\setenumerate{itemsep=0pt,topsep=2pt,parsep=0pt,partopsep=0pt,labelindent=3pt,leftmargin=*}
\usepackage{amsmath}
 
\usepackage{amsthm} 		
\theoremstyle{definition}

\usetikzlibrary{arrows, shapes.geometric, arrows.meta, shapes, decorations.pathreplacing, fit, patterns, patterns.meta}

\definecolor{Rcolor}{HTML}{E99595}
\definecolor{Gcolor}{HTML}{D9EAD3}
\definecolor{Bcolor}{HTML}{9DC3E6}
\definecolor{Ycolor}{HTML}{FFE699}
\definecolor{DGcolor}{rgb}{0.58, 0.77, 0.45}

\tikzstyle{expr} = [circle, minimum width=1.8cm, minimum height=1.8cm, text centered, align=center, inner sep=0, draw,font=\Large]
\tikzstyle{txt_huge} = [align=center, font=\Huge, scale=2]
\tikzstyle{txt} = [align=center, font=\Large]
\tikzstyle{cinn} = [double arrow, double arrow head extend=0cm, double arrow tip angle=130, shape border rotate=90, inner sep=0, align=center, minimum width=2.1cm, minimum height=2.3cm, fill=Bcolor, draw,font=\Large]
\tikzstyle{cinn_black} = [cinn, minimum height=2.5cm, fill=black]
\tikzstyle{arrow} = [thick,-{Latex[scale=1.0]}, line width=0.2mm, color=black]
\tikzstyle{xt} = [rectangle, align=center,  minimum width=3cm, minimum height=1.5cm,fill=Gcolor,font=\Large, rounded corners]

\definecolor{red_cb}{HTML}{e41a1c}
\definecolor{blue_cb}{HTML}{377eb8}
\definecolor{green_cb}{HTML}{4daf4a}
\definecolor{purple_cb}{HTML}{984ea3}
\definecolor{orange_cb}{HTML}{ff7f00}

\definecolor{EmeraldGreen}{HTML}{1ea78d}
\definecolor{EnglishRed}{HTML}{b02427}
\hypersetup{colorlinks=true,urlcolor=EmeraldGreen,citecolor=EmeraldGreen,linkcolor=EnglishRed}

\newcommand\one{\leavevmode\hbox{\small1\normalsize\kern-.33em1}}

\newcommand{\arXiv}[2][]{%
	\ifthenelse{\equal{#1}{}}%
	{\href{http://arxiv.org/abs/#2}{arXiv:#2}}%
	{\href{http://arxiv.org/abs/#2}{arXiv:#2~[#1]}}}

\def\slashchar#1{\setbox0=\hbox{$#1$}           
   \dimen0=\wd0                                 
   \setbox1=\hbox{/} \dimen1=\wd1               
   \ifdim\dimen0>\dimen1                        
      \rlap{\hbox to \dimen0{\hfil/\hfil}}      
      #1                                        
   \else                                        
      \rlap{\hbox to \dimen1{\hfil$#1$\hfil}}   
      /                                         
   \fi}

\newcommand{\tikznode}[2]{%
\ifmmode%
\tikz[remember picture,baseline=(#1.base),inner sep=0pt] \node (#1) {$#2$};%
\else
\tikz[remember picture,baseline=(#1.base),inner sep=0pt] \node (#1) {#2};%
\fi}

\def\mathswitchr#1{\relax\ifmmode{\mathrm{#1}}\else$\mathrm{#1}$\xspace\fi}
\def\mathswitch#1{\relax\ifmmode#1\else$#1$\xspace\fi}

\usepackage{layout}
\usepackage{colortbl}

\newcommand{\mjj}{\ensuremath{m_\mathrm{jj}}\xspace}
\newcommand{\mall}{\ensuremath{m_\text{all}}\xspace}
\newcommand{\mrsd}{\ensuremath{m_\text{RSD}}\xspace}
\newcommand{\GeV}{\giga\electronvolt}
\newcommand{\TeV}{\tera\electronvolt}
\newcommand{\gkk}{\ensuremath{G_\mathrm{KK}}\xspace}

\newcommand{\rtgg}{\ensuremath{R\to gg}\xspace}
\newcommand{\wpxy}{\ensuremath{W'\to XY}\xspace}

\newcommand{\Z}[2]{\ensuremath{Z_{#2}^{#1}}}
\newcommand{\Zshape}[2]{\ensuremath{Z_{#2,\mathrm{shape}}^{#1}}}
\hypersetup{
    pdftitle=Anomaly detection for multijet scenarios,
}

\begin{document}


\begin{center}{\Large \textbf{Anomaly detection for multijet scenarios
}}\end{center}
\begin{center}

Gregor Kasieczka\textsuperscript{1},  
Sung Hak Lim\textsuperscript{2,3},
Louis Moureaux\textsuperscript{1},
Tore von Schwartz\textsuperscript{1}, 
David Shih\textsuperscript{3}, and
Chitrakshee Yede\textsuperscript{1,$*$}

\end{center}

\begin{center}
{\bf 1} Institut für Experimental Physik, Universität Hamburg, Hamburg, Germany \\
{\bf 2} Particle Theory and Cosmology Group, Center for Theoretical Physics of the Universe,
Institute for Basic Science (IBS), 55 Expo-ro, Yuseong-gu, Daejeon, 34126, Republic of
Korea \\
{\bf 3} NHETC, Dept. of Physics and Astronomy, Rutgers University, Piscataway, NJ 08854, USA \\
* \href{mailto:chitrakshee.yede@cern.ch}{chitrakshee.yede@cern.ch}
\end{center}

\begin{center}
\today
\end{center}


\section*{Abstract}
{\bf
Signals of physics beyond the Standard Model continue to resist discovery at the LHC.
Recent years have seen the proliferation of new anomaly detection techniques, promising discovery with significantly fewer model assumptions than traditional approaches.
Strategies based on weak supervision have been especially successful, but so far were largely limited in scope by their reliance on a resonance manifesting a decay into a pair of jets.
In this work, we demonstrate that a well-established idea from jet substructure physics --- recursive soft drop --- in combination with the CATHODE technique for anomaly detection 
can be used to simultaneously perform anomaly detection for signals with an arbitrary number of jets in the final state, greatly increasing the scope of such searches.
}

\section{Introduction}

\noindent
Despite strong hints for its existence and an extensive search program, no physics beyond the Standard Model (BSM) has so far been observed at the LHC.
Anomaly detection --- see~\cite{Kasieczka:2021xcg, Aarrestad:2021oeb} for initial community overviews and~\cite{Belis_2024, Amram:2026vkc} for recent reviews --- has emerged as a new paradigm for searching in a less model-specific way than traditional searches. 
First experimental results by the CMS~\cite{CMS:2024nsz, CMS:2025sch} and ATLAS~\cite{ ATLAS:2020iwa, ATLAS:2023ixc, ATLAS:2023azi, ATLAS:2025obc} collaborations have further demonstrated the feasibility of this strategy.

Many anomaly detection searches are structured around detecting
signatures in which a resonance with specific decay products (e.g., two large-radius jets) is identified in the event.
This resonant assumption allows the construction of side-bands to estimate background and enables simple statistical analysis via bump-hunts.
However, it also limits the scope of the analyses as marginally more complicated final states (e.g., a three-jet resonance) would require a separate approach.

In this paper, we propose a method to search for a resonance decaying hadronically without restricting the number of jets in the final state.
This enables sensitivity to a wide class of signal models that cannot be found by existing anomaly detection techniques.
Our technique is based on the recursive soft-drop~\cite{Dreyer:2018tjj} algorithm, which we use to remove particles originating from soft and collinear radiation and recover the mass of the resonant particle.
This then either allows a bump-hunt search for resonances in the soft-drop mass distribution, either directly, or after further enhancement with the CATHODE~\cite{Hallin:2021wme,Hallin:2022eoq,Golling:2023yjq} technique.
We demonstrate the suitability of the recursive soft-drop mass and subsequent, weakly supervised enhancement of the anomaly on a number of simulated two-, three-, and four-prong signals. Furthermore, we investigate the impact of this approach on the Black Box 3 (BB3) dataset from the LHC Olympics~\cite{Kasieczka:2021xcg} challenge as a benchmark for our method.
BB3 contains a signal with two decay modes into different jet multiplicities, making it especially challenging with traditional approaches.

Other works have considered extending the scope of weakly supervised anomaly detection either by improving the input features, or the targeted topology. 
New sets of input observables are introduced in Refs.~\cite{Finke:2023ltw,Das:2026spe}. The studies consider a variety of features combining energy flow polynomials, high-level features, and sub-jettiness variables. The anomaly detector itself is made to select the most relevant inputs from the set. This approach resulted in more robust performance and consistently performed better than the curated feature sets used in previous studies. Using the full particle-level phase space is covered in Ref.~\cite{Buhmann:2023acn}. Anomaly detection using foundation models has been explored in Refs.~\cite{Mikuni:2024qsr, Mikuni:2025tar, Golling:2024abg, Li:2024htp, Bhimji:2025isp, Hsu:2026sww}.

Using CATHODE with event-level input features was proposed in Ref.~\cite{Brennan:2025fqy}.
The paper considers intermediate resonances produced in complex processes, with many particles present in the event alongside the resonance, which is assumed to decay to charged leptons (muons or taus).
The event discrimination is based on the simplex coordinates from Ref.~\cite{Cai:2024xnt}.
Compared to these results, we explore the case where the resonance has multiple decay modes.

Finally, extensions of CATHODE to further event topologies have been proposed.
Topologies involving missing transverse momentum were considered in Refs.~\cite{Finke:2022lsu,Kasieczka:2024lxf}.
Resonant and tail-based anomaly detection is combined in Ref.\cite{Bickendorf:2023nej} by applying the CATHODE method to search for supersymmetric gluino pair production.
CATHODE was also applied to soft-unclustered-energy patterns in Ref.~\cite{Curtin:2025ksm}.
Additional studies have been performed to rediscover $\Upsilon$ meson to dimuons using CMS open data in Ref.~\cite{Gambhir:2025afb}, and the advances in the Higgs$+X$ anomaly detection (HAXAD) strategy have been presented in Ref.~\cite{cheng2026anomalydetectionsearchesnew}. 
However, none of these have demonstrated simultaneous sensitivity to multiple decay modes as presented here.

This paper is structured as follows. In section~\ref{sec:dataset} we introduce all the datasets and signal topologies used in this paper. Section~\ref{sec:resonant_feat} discusses in detail the challenge in choosing the resonant feature and shows supporting studies. Section~\ref{sec:ADsetup} details our anomaly detection setup introducing the input feature sets and the models used. The performance for different signal topologies and results from additional studies are discussed in section~\ref{sec:results}. We summarize in section~\ref{sec:summary}. 

\section{Datasets}
\label{sec:dataset}

We generate dijet and four-jet events with \wpxy with 
$m_{W'}=\SI{4.2}{\TeV}$ and $X$ and $Y$ each decaying to a pair of light quarks,
a signal inspired by the LHC Olympics R\&D dataset.
The dijet events are generated with $m_X=500$ and $m_Y=\SI{100}{\GeV}$.
For four-jets events, we consider a case with boosted $X$ and $Y$ by setting $m_X=m_Y=\SI{1}{\TeV}$, an imbalanced case with $m_X=1$ and $m_Y=\SI{3}{\TeV}$, and a balanced case with $m_X=m_Y=\SI{2}{\TeV}$.
We generate \num{10000} events for each signal, with a trigger efficiency between \num{50} and \SI{80}{\percent}.

We also consider the signal events from the LHC Olympics~\cite{Kasieczka:2021xcg} Black Box 3 dataset.
The signal events in this dataset correspond to a Kaluza-Klein gluon \gkk at \SI{4.2}{\TeV} with two simulated decay modes: direct decay to a pair of light quarks and decay through an intermediate radion \rtgg with $m_R=\SI{2.217}{\TeV}$.
There are \num{1200} events of the dijet (first) decay mode and \num{2000} events of the trijet (second) mode, which we increase to \num{15000} of each.
We obtain an intermediate case by setting the radion mass to $m_R=\SI{500}{\GeV}$.
To increase event complexity, we also generate samples at the same mass points with $R\to t\bar t\to\text{hadrons}$.

For background events, our starting point is also the BB3 dataset with 1 million QCD dijet background examples.
We used all the background in the dataset for training our models and additionally, generated 2 million background events in the same format for evaluation.

All events are simulated with \textsc{Pythia}~8.219~\cite{Sjostrand:2006za,Sjostrand:2014zea} and \textsc{Delphes}~3.4.1~\cite{deFavereau:2013fsa,Mertens:2015kba}. The events are selected using a \SI{1.2}{\TeV} jet trigger based on the anti-$k_\mathrm T$ algorithm~\cite{Cacciari:2008gp} with distance parameter $R=1$.

We also use the LHC Olympics R\&D dataset for cross-checks.
It contains one million background and \num{10000} signal events generated in the
same way as Black Box 3, but with different Pythia settings~\cite{Kasieczka:2021xcg}.
The signal process is \wpxy with $X, Y\to q\bar q$ and $m_{W'}=\SI{3.5}{\TeV}$,
$m_X=\SI{500}{\GeV}$, and $m_Y=\SI{100}{\GeV}$.
Similar to Black Box 3, we generate an additional 2 million background events
for use in our studies.

\section{Resonant Feature}
\label{sec:resonant_feat}

\subsection{Resonant Feature Definition}

\begin{figure}[t]
    \centering
    \includegraphics[width=\linewidth]{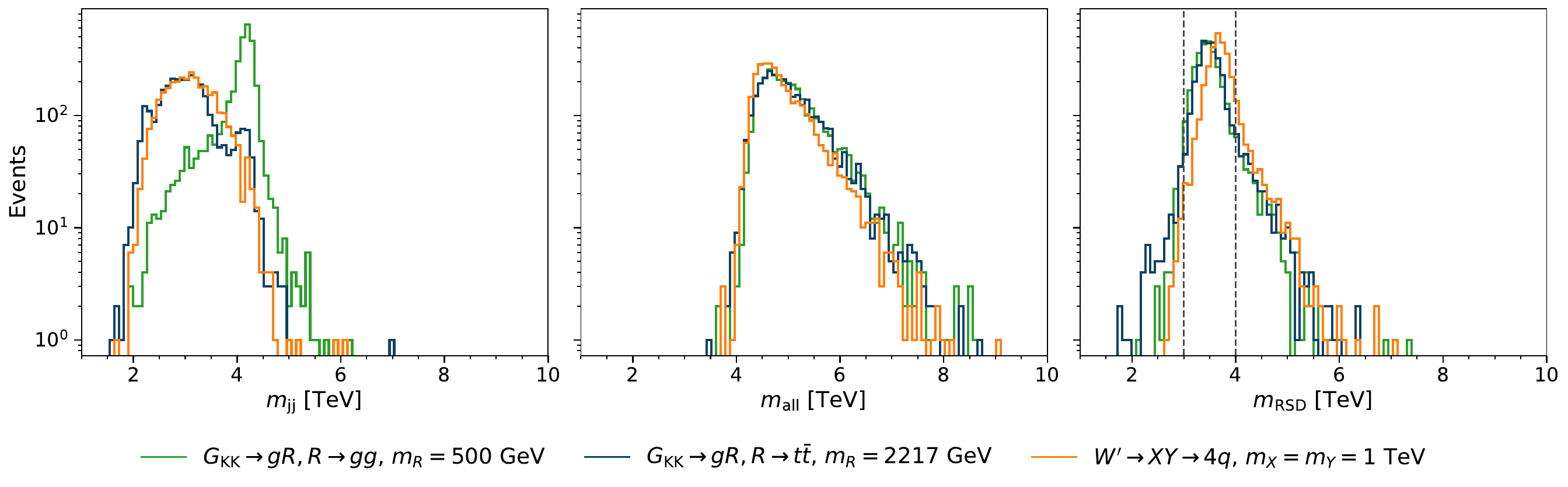}
    \caption{The three mass distributions for two-, three-, and four-prong signals.
    Shown are the dijet mass ($m_{\textrm{jj}}$, left), the mass of all jets in the event ($m_{\textrm{all}}$, center), and the recursive soft-drop mass ($m_{\textrm{RSD}}$, right)
    }
    \label{fig:masses}
\end{figure}

Resonant anomaly detection in general, and the CATHODE method in particular, are designed and optimized around the presence of a resonant feature. For the dijet case, the dijet invariant mass, \mjj provides a well defined resonant feature. However, this obviously fails for multijet scenarios.
This behavior is illustrated in Fig.~\ref{fig:masses} (left). It shows that a dijet resonance at 4.2~TeV (green) is indeed well reconstructed, while the tri- (dark blue) and four-jet (orange) resonances are not fully reconstructed and much broader.
In order to be more generic to signals involving multiple jets, we investigate the use of event-level variables.

A first step to reconstruct resonances with an arbitrary number of jets in the final state would be to construct \mall, the invariant mass of all reconstructed particles in an event.
The resulting distributions are displayed in Fig.~\ref{fig:masses} (center).
We observe that indeed now all three distributions overlap.
However, the \mall definition leads to a very broad signal distribution with no sharp peak and a pronounced tail at higher mass, losing the resonant structure.
This is caused by particles not originating from the hard process contributing to \mall, such as initial-state radiation and soft particles from the underlying event.

We thus investigate an algorithm designed to remove these contributions, namely recursive soft drop.
The recursive soft-drop (RSD)~\cite{Dreyer:2018tjj} algorithm is a jet substructure technique designed to groom jets by removing
soft and collinear radiation.
It considers the clustering history of the Cambridge-Aachen algorithm~\cite{Dokshitzer:1997in,Wobisch:1998wt} in reverse order.
The lowest-$p_\mathrm T$ subjet of each clustering step is removed until the following condition is met:
\begin{equation}
    \frac{\min(p_{\mathrm T,1}, p_{\mathrm T,2})}{p_{\mathrm T,1} + p_{\mathrm T,2}}
    >
    z_\text{cut} \left( \frac{\Delta R_{12}}{R_0} \right)^\beta,
\end{equation}
where $p_{\mathrm T,i}$ are the subjet's transverse momenta, $\Delta R_{12}=\sqrt{(\phi_2-\phi_1)^2 + (\eta_2-\eta_1)^2}$ is the distance in azimuth and pseudorapidity, $z_\text{cut}$ and $\beta$ are parameters, 
and $R_0$ is a dynamical normalization scale set to the angle $\Delta R_{12}$ at each step~\cite{Dreyer:2018tjj}, initially starting at 1.

We apply RSD to all particles in every event, considering it as a single jet, and reconstruct the invariant mass of the remaining constituents, which we define as \mrsd. Optimizing the RSD parameters for each signal hypothesis we consider would lead to a total loss of model-agnosticity, but they do have to be chosen somehow. We opt to tune $z_\text{cut}$ and $\beta$ on two signal processes, selecting values that maximize the overlap between the mass distributions of a $G_\mathrm{KK}$ and a $W'$ signal, and then test whether this choice is effective for the other signals in our study. For a given $z_\text{cut}$ and $\beta$, we compute the 16th, 50th and 84th percentiles of the signal \mrsd distributions.
We then select the set that minimizes the differences between corresponding percentiles, leading to maximum overlap between the mass distributions.
We obtain $z_\text{cut}=0.15$ and $\beta=0.7$.
These values lead to consistent performance across all the signals we consider.

\begin{figure}[t]
    \centering
    \includegraphics[width=0.45\linewidth]{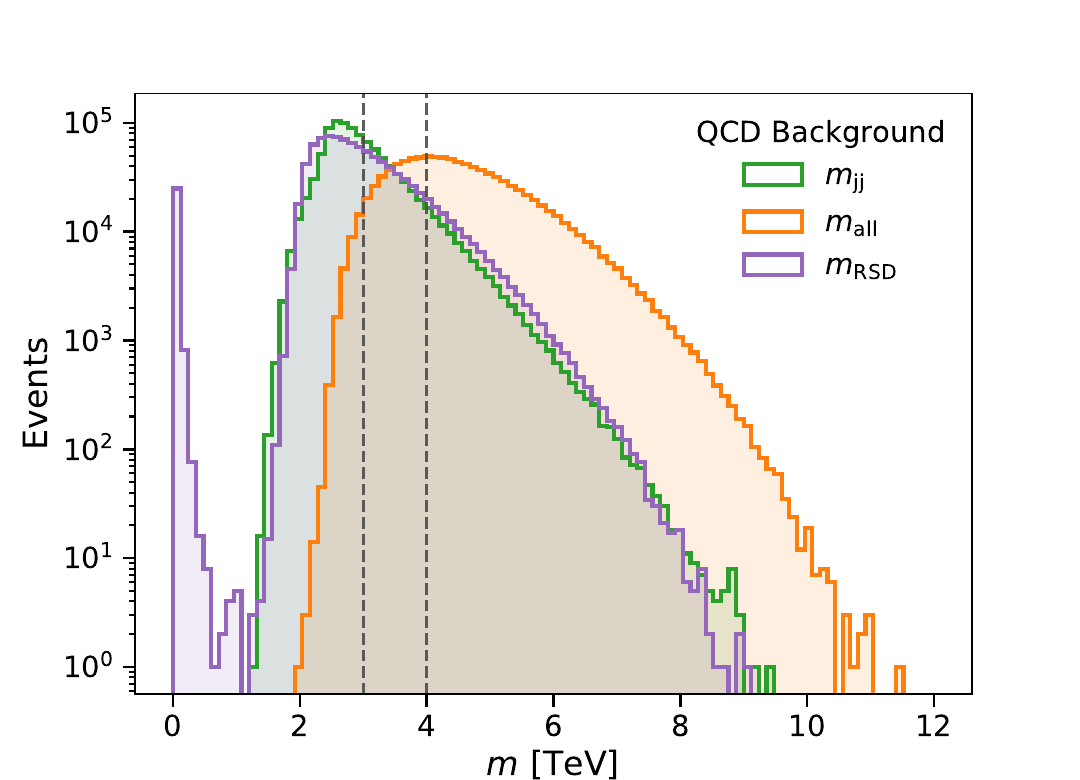}
    \includegraphics[width=0.45\linewidth]{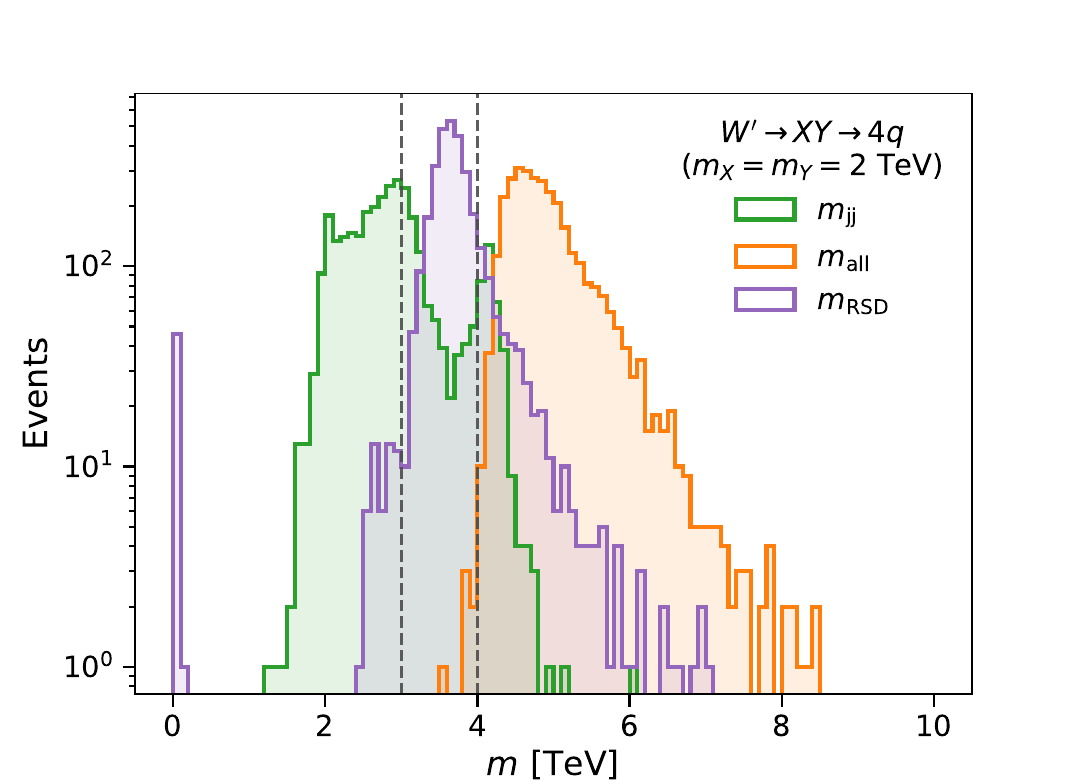}
    \caption{%
        Comparison of the mass definitions for the QCD background and a four-jet signal. The \mrsd signal region is marked with the vertical dotted lines.}
    \label{fig:WXYto4q_allMass}
\end{figure}

We observe that the chosen RSD settings lead to signal distributions peaking around \SI{3.5}{\TeV} as shown in Fig.~\ref{fig:masses} (right).
The signal peaks also overlap as intended, and a signal region (SR) from \num{3} to \SI{4}{\TeV}, also shown on the plots, 
captures about 80\% of the signal events irrespective of the decay topology.
The shift in the mass distributions, with the peaks appearing below the corresponding resonance masses, is not an issue
for discovery.
A calibrated mass measurement can always be performed afterward with other techniques.

The resulting mass distributions for the background and a four-jets signal are shown in Fig.~\ref{fig:WXYto4q_allMass}. Within the marked signal region, the background remains steeply falling and signal exhibits a peak. 

\subsection{Impact on Sensitivity}

We calculate the statistical significance $Z$ using the numbers $S$ ($B$) of signal (background) events, as $Z = S/\sqrt B$. 

To further illustrate the impact of the choice of a resonant feature,  we extract the expected significance from the
\mjj, \mall, and \mrsd distributions for an injection of \num{3200} signal events and \num{996800} background events for each considered signal model. This is the same amount of signal as used in the original LHC Olympics study and results in the inclusive initial significance (\Z{inc}{0}) of \num{3.2}.

We first consider simple counting experiments (i.e. no machine learning to enhance the signal fraction), for which we define a signal region
around resonant peaks.

None of the considered signal models is resonant in \mall, and many of them do not
exhibit a resonant peak in \mjj; this is illustrated in Fig.~\ref{fig:WXYto4q_allMass} 
with the four-jets signal.
We choose $\num{4}<\mjj<\SI{4.4}{\TeV}$, $\num{4}<\mall<\SI{5}{\TeV}$, and
$\num{3}<\mrsd<\SI{4}{\TeV}$ for the SR boundaries.
The results are shown in the first part of Table~\ref{tab:significance-combine}.
Results marked with an asterisk ($*$) are cases where signals are not resonant in the respective variable.

While \mjj works well for dijet signals, it fails for other topologies.
On the other hand, \mrsd works equally well for all signal topologies.
For the original Black Box 3 signal mix,
the \mrsd signal region contains about \num{2700} signal and \num{305000} background events,
corresponding to a initial significance in the SR (\Z{\mrsd}{0}) slightly below $5\sigma$.
We note that in this case, using the \mjj-based SR also leads to an initial significance (\Z{\mjj}{0}) of $4.4\sigma$, which was not reported by 
any team during the LHC Olympics community challenge~\cite{Kasieczka:2021xcg}.
To our knowledge, the best result on Black Box 3 in the literature was obtained in Ref.~\cite{Matos:2024ggs}
with the ANTELOPE method, reporting an improvement of about 1.4 times the initial inclusive significance of $3.2\sigma$,
for a final significance of $4.5\sigma$---similar to our result with \mjj 
and slightly worse than the improvement achieved by \mrsd.

Since the signal distributions in these variables are not necessarily resonant, making this approach limited, we use
an alternative approach based on binned maximum likelihood fits of the background and signal shapes, implemented in 
\textsc{Combine}~\cite{CMS:2024onh} as described in Appendix~\ref{sec:combine}.
This method exploits the exact mass distribution of the signal events (assuming it is known) to constrain the signal
yield, which can lead to greater sensitivity, and we use it as an upper bound.
The expected significances (\Zshape{m}{0}) obtained using this method are shown in the last three columns of Table~\ref{tab:significance-combine}.
We again find that, while \mjj is the best variable for dijet signals, it is not as sensitive to other topologies.
Fits based on \mall and \mrsd yield inclusive significances around 3 and $5\sigma$, respectively,
consistently for all signal models.

\begin{table}[h!]
    \centering
    \caption{%
        Significance, $S/\sqrt B$, in the SR (\Z{m}{0}) and the full inclusive mass spectrum using 
        binned-likelihood fits (\Zshape{m}{0}), based on $m = \mjj$, \mall,
        and \mrsd distributions for \num{3200} events of the different signals.
        The best-performing variable is indicated in bold for each significance
        definition. Results marked with an asterisk ($*$) for \Z{m}{0} are cases where signals are not resonant in the respective variable.
     }
    \label{tab:significance-combine}
    \begin{tabular}{clcccccc}
        \toprule
        \#jets & Signal process & \Z{\mjj}{0} & \Z{\mall}{0} & \Z{\mrsd}{0} & \Zshape{\mjj}{0} & \Zshape{\mall}{0} & \Zshape{\mrsd}{0} \\
        \midrule
        2 & $G_\mathrm{KK}\to q\bar q$
            & \bf 10 & 2.8* & 4.8
            & \bf 10 & 3.1 & 5.7 \\
        2 & $G_\mathrm{KK}\to gR$, $R\to gg$ ($m_R=\SI{500}{\GeV}$)
            & \bf 9.5 & 2.7* & 5.0
            & \bf 9.7 & 3.1 & 4.8 \\
        2 & $G_\mathrm{KK}\to gR$, $R\to t\bar t$ ($m_R=\SI{500}{\GeV}$)
            & \bf 9.2 & 2.9* & 4.9
            & \bf 9.3 & 3.1 & 4.9\\
        3 & $G_\mathrm{KK}\to gR$, $R\to gg$ ($m_R=\SI{2217}{\GeV}$)
            & 0.9* & 2.8* & \bf 5.1
            & 1.2 & 3.1 & \bf 4.6 \\
        3 & $G_\mathrm{KK}\to gR$, $R\to t\bar t$ ($m_R=\SI{2217}{\GeV}$)
            & 1.1* & 2.9* & \bf 4.9
            & 1.3 & 3.0 & \bf 4.9 \\
        2 & \makecell[l]{$W'\to XY\to 4q$ \\ ($m_X=\SI{500}{\GeV}$,$m_Y=\SI{100}{\GeV}$)}
            & \bf 11 & 3.4* & 4.5
            & \bf 11 & 2.9 & 6.1 \\
        4 & $W'\to XY\to 4q$ ($m_X=m_Y=\SI{1}{\TeV}$)
            & 0.6* & 3.4* & \bf 4.8
            & 1.4 & 3.3 & \bf 5.9 \\
        4 & $W'\to XY\to 4q$ ($m_X=m_Y=\SI{2}{\TeV}$)
            & 1.5* & 3.4* & \bf 4.7
            & 1.5 & 3.3 & \bf 5.8 \\
        4 & $W'\to XY\to 4q$ ($m_X=\SI{1}{\TeV}$,
                          $m_Y=\SI{3}{\TeV}$)
            & 0.8* & 3.4* & \bf 4.8
            & 2.5 & 3.3 & \bf 5.4 \\
        \addlinespace
        $2+3$ & Black Box 3 mix
            & 4.4 & 2.8* & \bf 4.9
            & 4.1 & 3.1 & \bf 4.6 \\

        \bottomrule
    \end{tabular}
\end{table}

\begin{figure}
    \centering
    \includegraphics[width=0.6\linewidth]{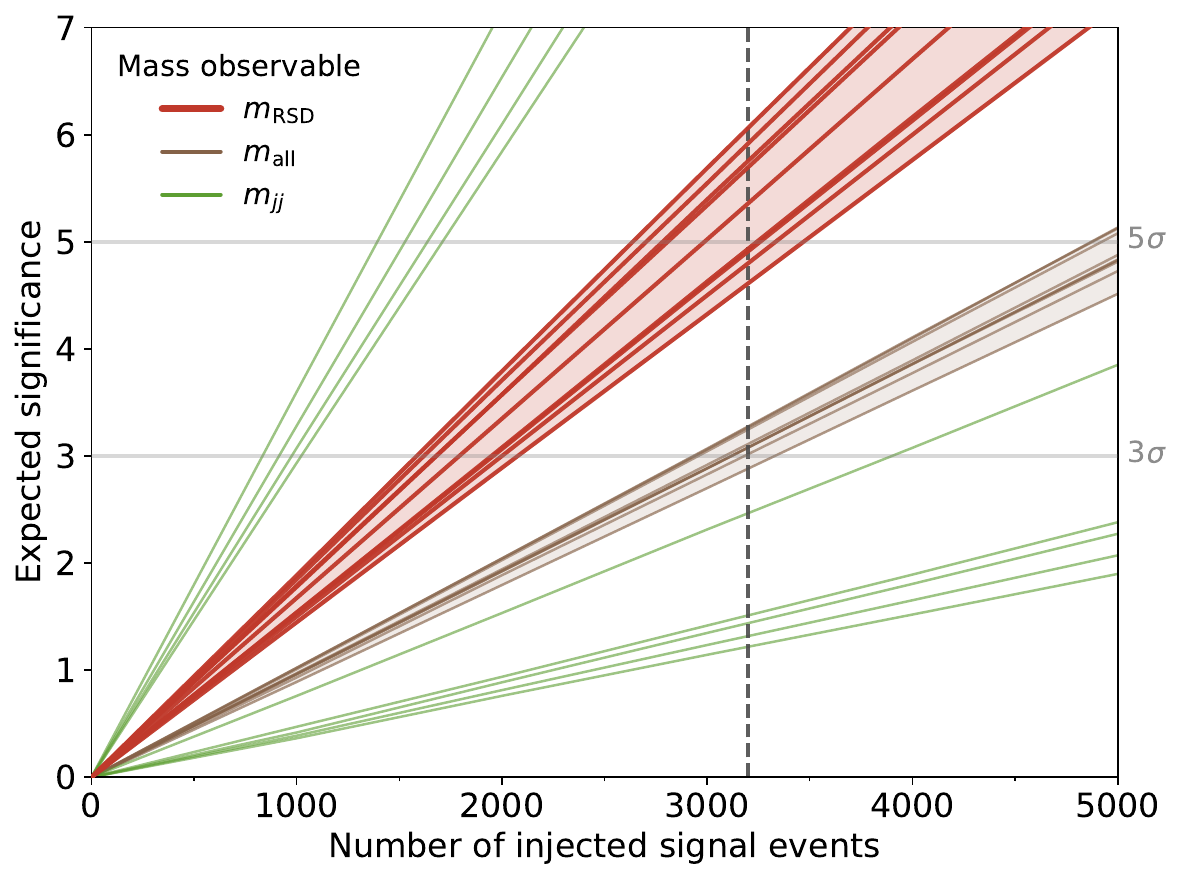}
    \caption{Expected significance from shape-based fits on the full inclusive mass spectrum shown for different signal injections for all signals grouped by color according to the mass definitions. Those defined with \mjj split into two clear groups: dijet (faster turn-on) and multijet (slower turn-on).
    }
    \label{fig:exp_sig}
\end{figure}

For the shape-based significance, we also report the number of events needed to reach the symbolic thresholds for observation, $Z=3$. The significance increases linearly with increasing numbers of signal events, as shown in Fig.~\ref{fig:exp_sig}.
The figure shows that smaller signal injections lead to lower significance and a relatively large signal yield is required to reach a substantial significance---around 1700 signal events for observation and 3200 events for discovery.
In realistic scenarios, since such high signal yield is not expected, it motivates the use of resonant anomaly detection methods to further enhance the sensitivity.

\section{Anomaly Detection Setup}
\label{sec:ADsetup}

We now investigate the use of anomaly detection to further enhance the significance.
The idea is to train a classifier to distinguish between signal and background based on a set of \emph{auxiliary features}, allowing us to reject background and enhance the significance of the signal. By training the classifier in a specific way, it can learn the signal properties directly from the data, leaving the analysis model-agnostic.
We use the CATHODE method~\cite{Hallin:2021wme} for this purpose and compare it with an Idealized Anomaly Detector (IAD), described in Section~\ref{sec:clsfs}.

To perform the CATHODE method, the data is split into a signal region (SR) and side-bands (SB) based on a feature in which the signal is expected to be resonant, in our case \mrsd. A generative model is trained in the SB to learn the conditional density of auxiliary features with respect to the mass. A classifier is then trained to compare the background pseudo-events sampled from the generative model and the SR data. This weakly supervised classifier trained to distinguish between data and background eventually learns the signal-to-background likelihood ratio.
We provide more details about our implementation of CATHODE in the following sections.

\subsection{Input Features}
\label{sec:features}

\paragraph{Event-level feature set}
In keeping with our objective of detecting signals regardless of the jet multiplicity,
we consider a feature set made of event-level variables. 
We use the scalar sum of all transverse momenta, $H_\mathrm T$,
and the first four jettiness~\cite{Stewart:2010tn} variables $\tau_1$ to $\tau_4$
as input features for classification.
For all jettiness calculations, we adopt the \texttt{OnePass\_WTA\_KT\_Axes} with the
unnormalized measure~\cite{Thaler_2012} as implemented in 
\textsc{FastJet}~3.4.3~\cite{Cacciari:2011ma}, as this is robust against soft radiation.
We do not use recursive soft drop for these features, as this led to reduced performance. This constitutes the main feature set used in this paper.
The distribution of the features is shown in Fig.~\ref{fig:bkgModeling} for background and a few example signal samples along with background sampled from the generative model.

\paragraph{Dijet feature set}
For additional studies, we also use the original CATHODE feature set~\cite{Hallin:2021wme}:
the invariant mass of the lighter jet ($m_{J1}$), the difference between the invariant mass 
of two leading jets ($\Delta m_{JJ} $), and the subjettiness ratios $\tau_{21}=\tau_2/\tau_1$
for the two jets ($\tau_{21}^{J1}$ and $\tau_{21}^{J2}$)~\cite{Thaler_2011,Thaler_2012}.
The jets are clustered with anti-$k_\mathrm T$ algorithm with radius parameter $R=1$.
For consistency, we use the same axes definition and measure as in the event-level feature set,
as opposed to the \texttt{OnePass\_KT\_Axes} and normalized measure used in the CATHODE paper.
This leads to a slight reduction in performance as compared to the original implementation. 

\subsection{Generative Model}

The generative model used in this paper to interpolate the background density into the signal 
region uses conditional flow matching~\cite{lipman2023flowmatchinggenerativemodeling}
with four transformer~\cite{vaswani2023attentionneed} blocks.
This more powerful architecture, described below, was chosen over the normalizing flows used in CATHODE for its better density estimation and interpolation performance.
The model is conditioned on the resonant variable, $m=\mrsd$ or \mjj.
It is first trained in the SB, using all the available pseudo-data. Background-like events are then sampled from the trained model, with the conditional variable $m \in \mathrm{SR}$. 

While training, we use background events from the LHCO datasets, split into training (\SI{90}{\percent}) and validation (\SI{10}{\percent}). We discard the events peaking around a mass of zero. 
The input auxiliary features were first scaled to the interval $(0,1)$ and then logit-transformed, followed by standardization using \texttt{StandardScaler}. The mass feature was standardized using \texttt{StandardScaler}. These transformations improve the accuracy of the density estimator as studied in the CATHODE paper. Background-like events are then sampled from the trained model using $m \in \mathrm{SR}$ sampled from a kernel density estimator (KDE) fitted to the $m$ values of the SR training set. The KDE is implemented using scikit-learn~\cite{JMLR:v12:pedregosa11a} with a Gaussian kernel with the bandwidth of 0.01.
We generate as many background samples as there are events in the SR.
They provide the background estimate used to train the CATHODE classifier in the next step.

Each transformer block contains a multihead attention~\cite{vaswani2023attentionneed} layer 
followed by the layer normalization, a feed-forward module that contains two linear layers with 
GELU~\cite{hendrycks2023gaussianerrorlinearunits} activation and dimension 256, and another
layer normalization. A dropout of 0.1 is applied after each linear layer and these layers operate via residual connections. 
Training is done for \num{30000} epochs using large batch size of $2^{14}$, optimized using an initial learning rate of \num{1e-3} that goes to \num{1e-6} with a cosine annealing~\cite{loshchilov2017sgdrstochasticgradientdescent} scheduler for smooth convergence. The scheduler's T\_max parameter was set to the total number of training epochs.

\begin{figure}
    \centering
    \includegraphics[width=0.5\textwidth]{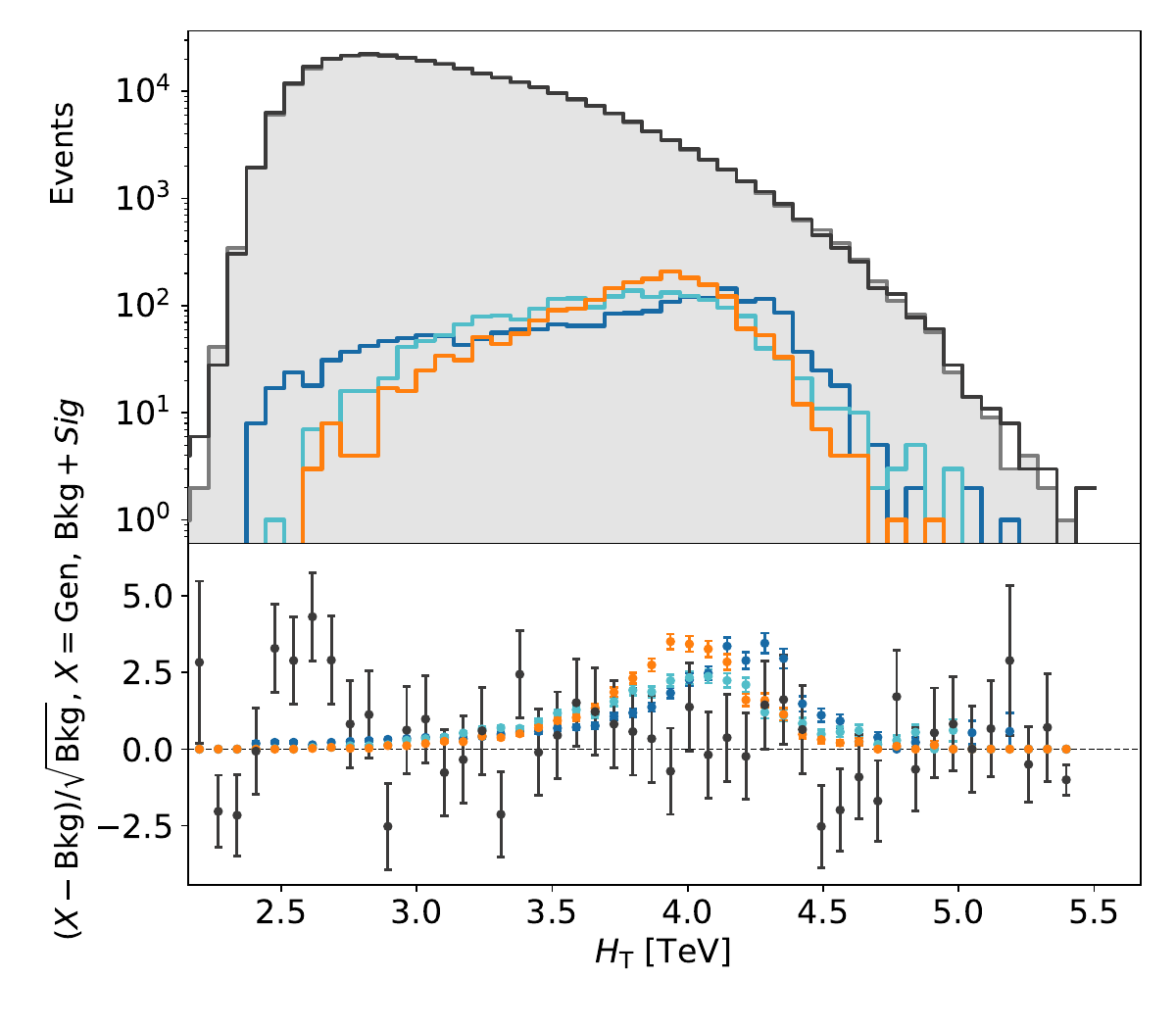}\relax
    \includegraphics[width=0.5\textwidth]{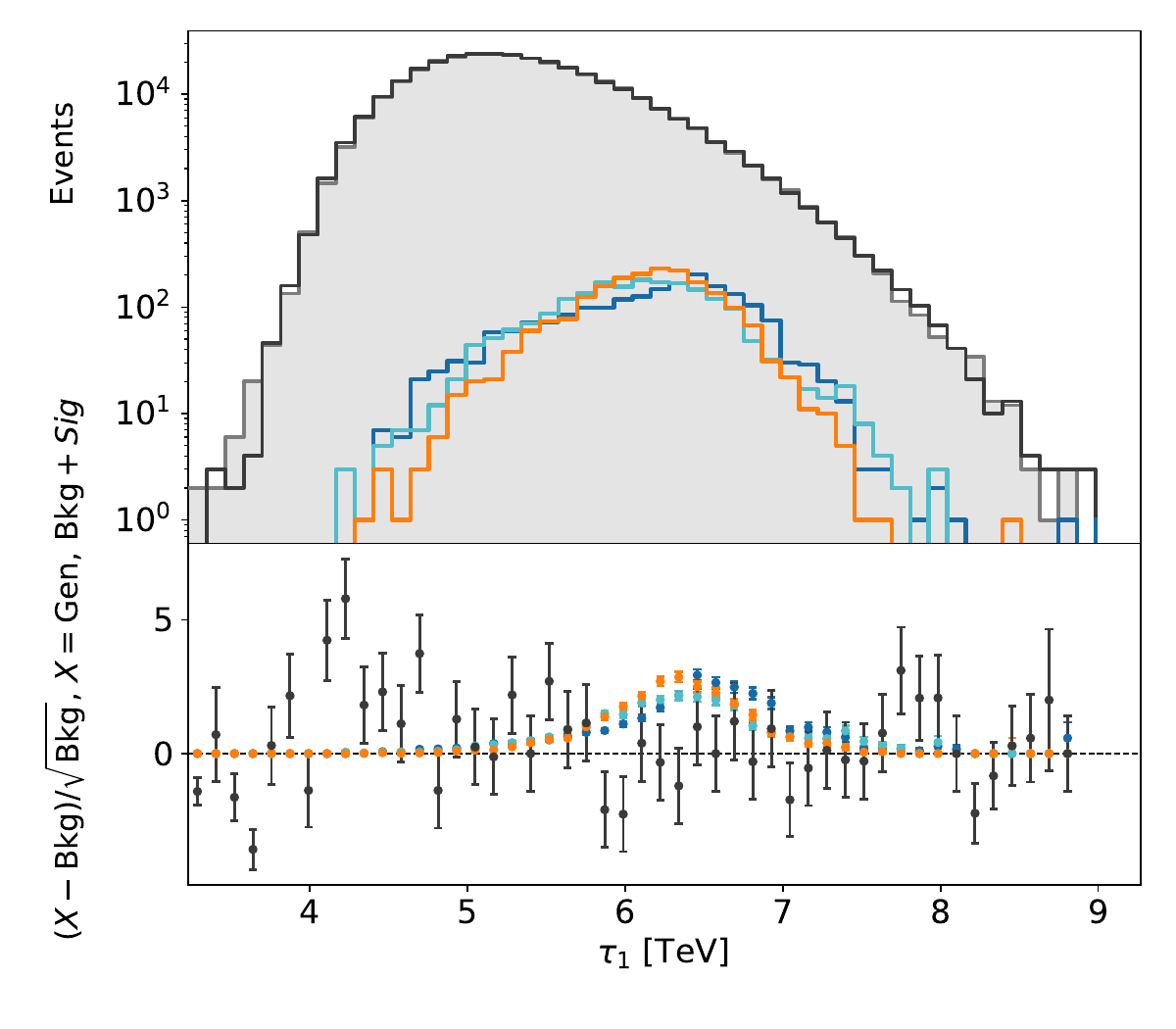}
    \includegraphics[width=0.5\textwidth]{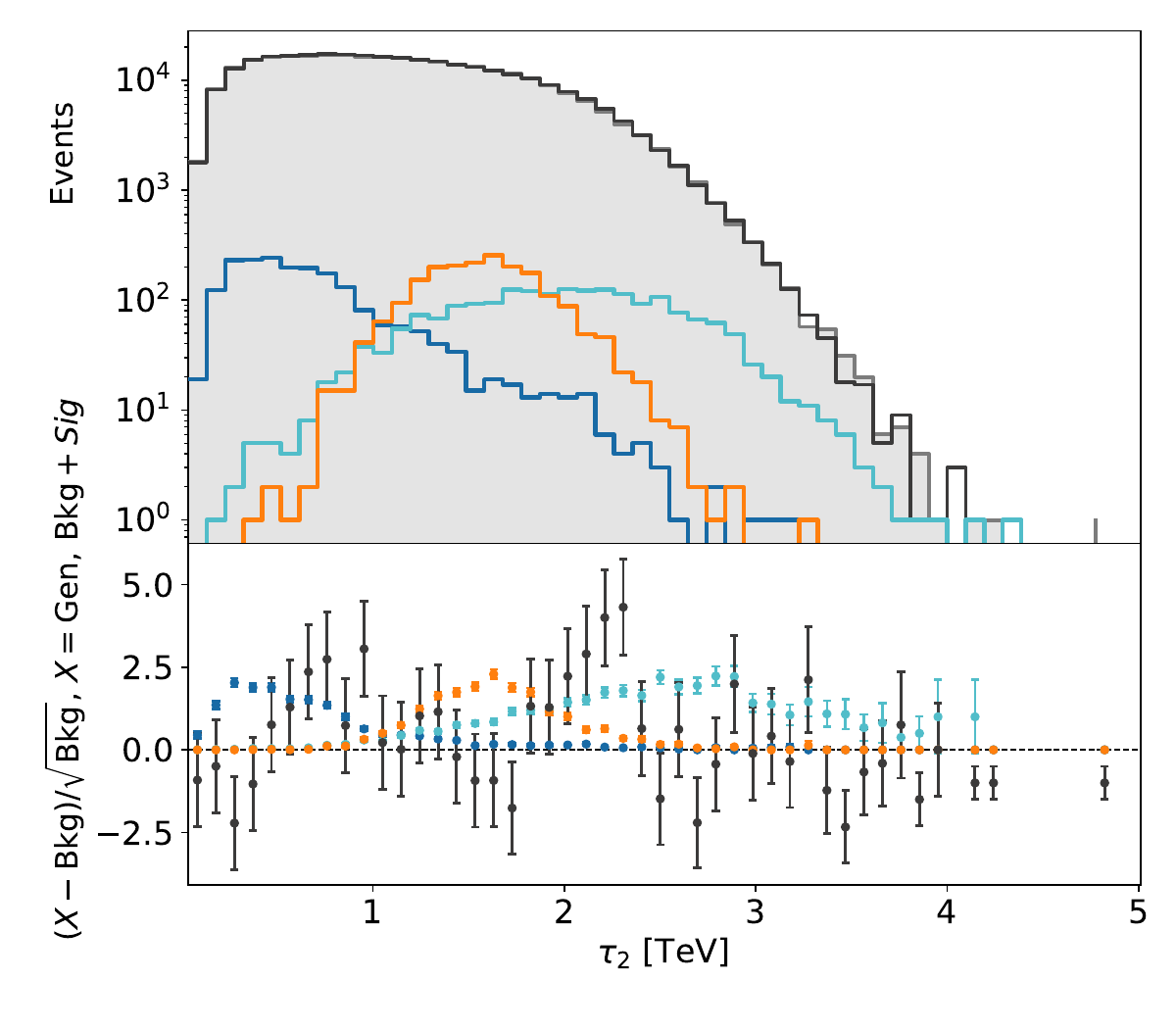}\relax
    \includegraphics[width=0.5\textwidth]{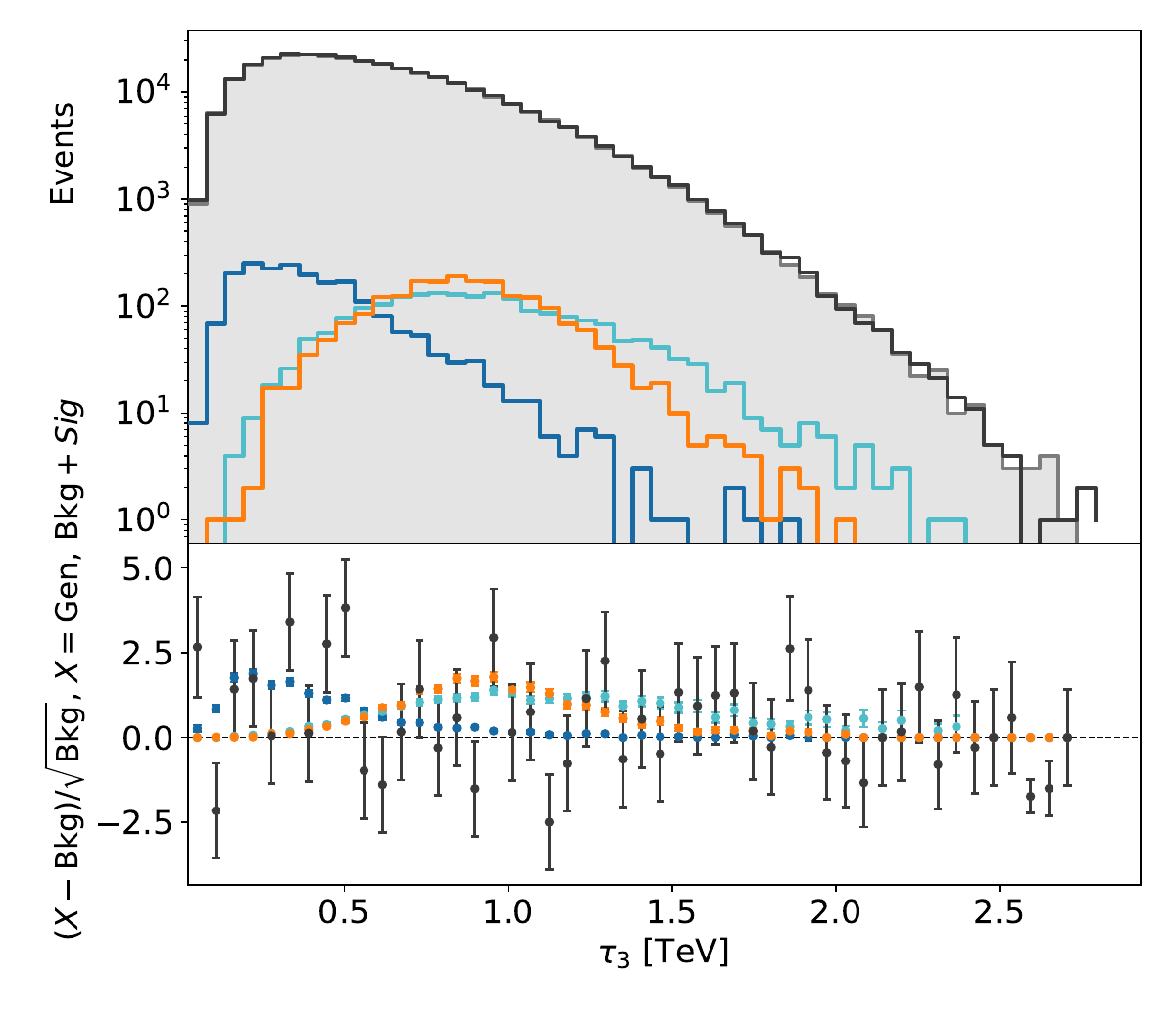}
    \includegraphics[width=0.5\textwidth]{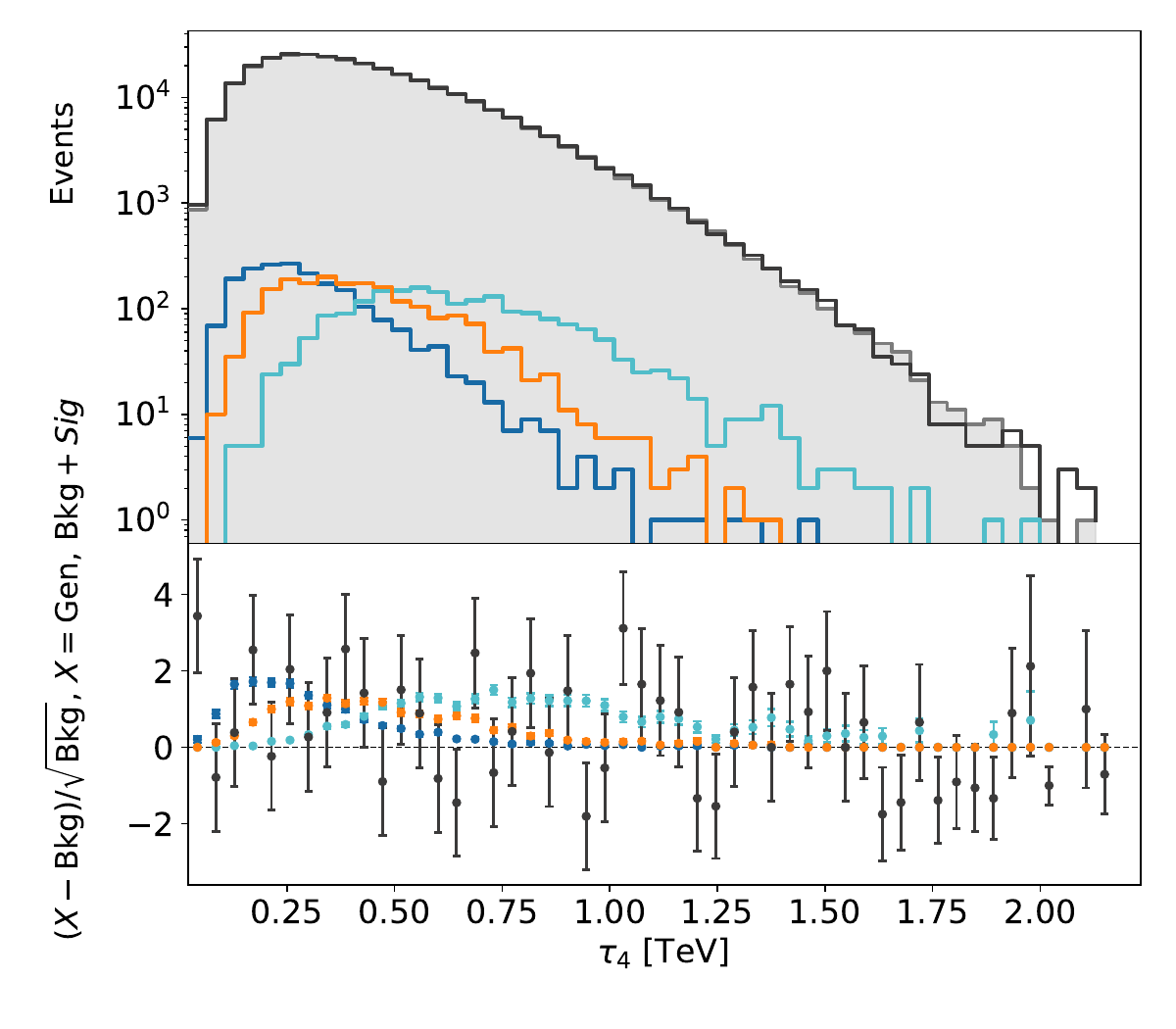}\relax
    \includegraphics[width=0.5\textwidth]{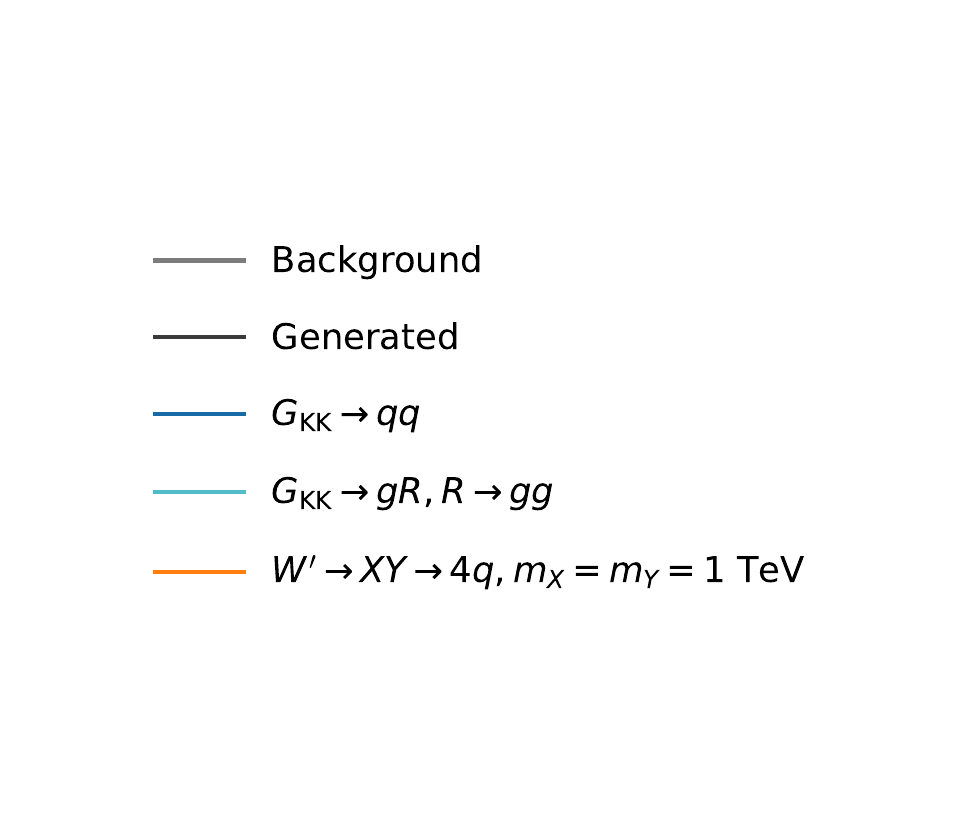}

    \caption{%
    Distributions comparing the background in the signal region to the generated background for the five event-level auxiliary features used with 2000 signal events, in the signal region ($3<\mrsd<\SI{4}{\TeV}$). 
    The ratio panel compares the pulls of the signals and the generated background. 
    }
\label{fig:bkgModeling}
\end{figure}

The background modeling is shown in Fig.~\ref{fig:bkgModeling}.
The generated samples closely follow the true background distribution.

\subsection{Classifiers}
\label{sec:clsfs}
While the generative model provides a good description of the background, it is not
perfect and remaining background mismodelling can affect the classifier performance.
We thus consider two kinds of classifiers in the SR:

\begin{itemize}
\item\textbf{Idealized anomaly detector (IAD)}\quad
This classifier represents the performance of our meth\-od if we could have a perfect
background estimate, providing an upper bound on anomaly detection performance.
It is trained using background events drawn from the extra generated dataset, with label 0,
and SR data with label 1.

\item\textbf{CATHODE classifier}\quad
The CATHODE classifier corresponds to the full anomaly detection pipeline.
It is trained using background pseudo-events sampled from the generative model,
with label 0, and SR data with label 1.
\end{itemize}

The classifiers are fully connected neural networks implemented in PyTorch~\cite{paszke2019pytorchimperativestylehighperformance},
with three hidden layers of 64 nodes each.
Hidden layers use ReLU activation~\cite{Nair2010ReLU} and the output node uses sigmoid activation.
The networks are trained using the Adam~\cite{kingma2017adammethodstochasticoptimization}
optimizer with a learning rate of 0.001 until the validation loss does not improve for 10 consecutive epochs.
The binary cross-entropy loss~\cite{mao2023crossentropylossfunctionstheoretical} is used for training.
While training an IAD classifier, the batch size is set to 1024 to avoid model collapse. 
Unlike the original CATHODE implementation, no oversampling from the generative model is employed.
No benefit from oversampling was observed, which we attribute to the large number of events present in the signal region. All classifiers are evaluated on the first same signal events kept aside for evaluation. For the IAD, we use the next $n$ events for training and validation depending on the injection.
 
Both classifiers share the same architecture and are trained and evaluated on the datasets of the same size.
We use the pseudo-data background events from the LHCO datasets split into training (\SI{90}{\percent}) and validation (\SI{10}{\percent}) that are in SR, to which various numbers of signal events are added in the same ratio.
All classifiers are then evaluated on a held-out set of \num{1000} signal events and one million background events generated additionally.

\section{Results}
\label{sec:results}

We show results for a single decay channel, using each signal individually. We further demonstrate the case for mixed decays where signal is a mixture of two decay channels, treated collectively as one signal class. We later show another study of the performance using the R\&D dataset comparing dijet feature sets to our event-level feature set.

We choose to show results at a background efficiency of \SI{1}{\percent} for all results.
The results are stable with respect to this choice and the significance changes by less
than \SI{30}{\percent} when using \SI{0.1}{\percent} instead.
We repeat each training ten times with different random network initializations and show the median value.

\subsection{Single Decay Channel}

We start by evaluating the sensitivity of the IAD and CATHODE with the event-level
feature set in the \mrsd SR.
We inject a variable number of events from all benchmark models and for each injection, train 10 classifiers to compute the median significance after selecting events.
For CATHODE, we also show the fluctuations in the classifier training with error bars that span the interval between the quantiles at \num{16} and \SI{84}{\percent}. 

\begin{figure}
    \includegraphics[width=0.5\textwidth]{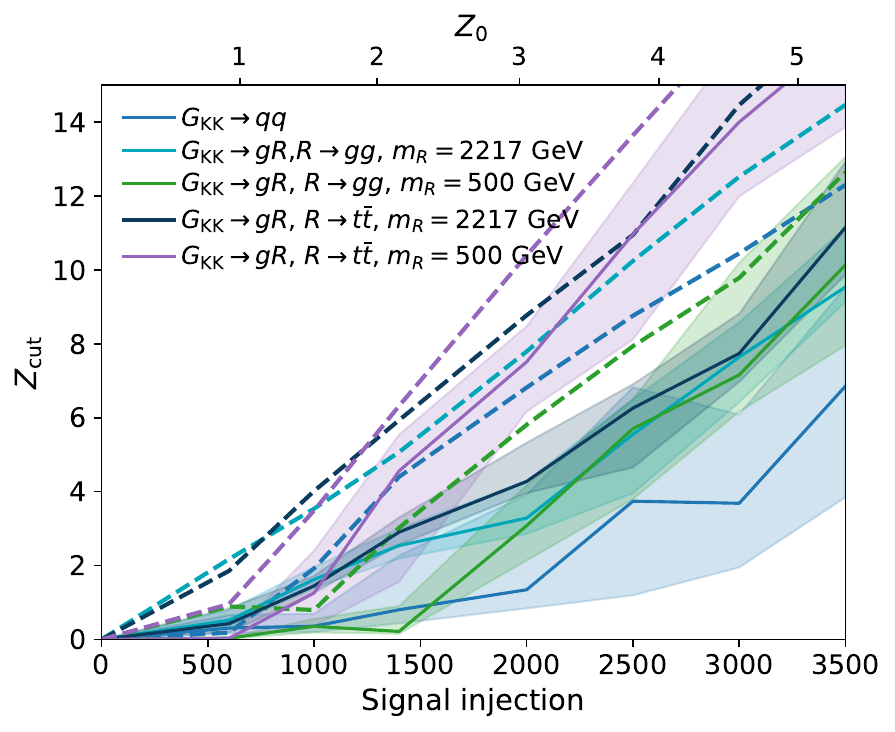}\relax
    \includegraphics[width=0.5\textwidth]{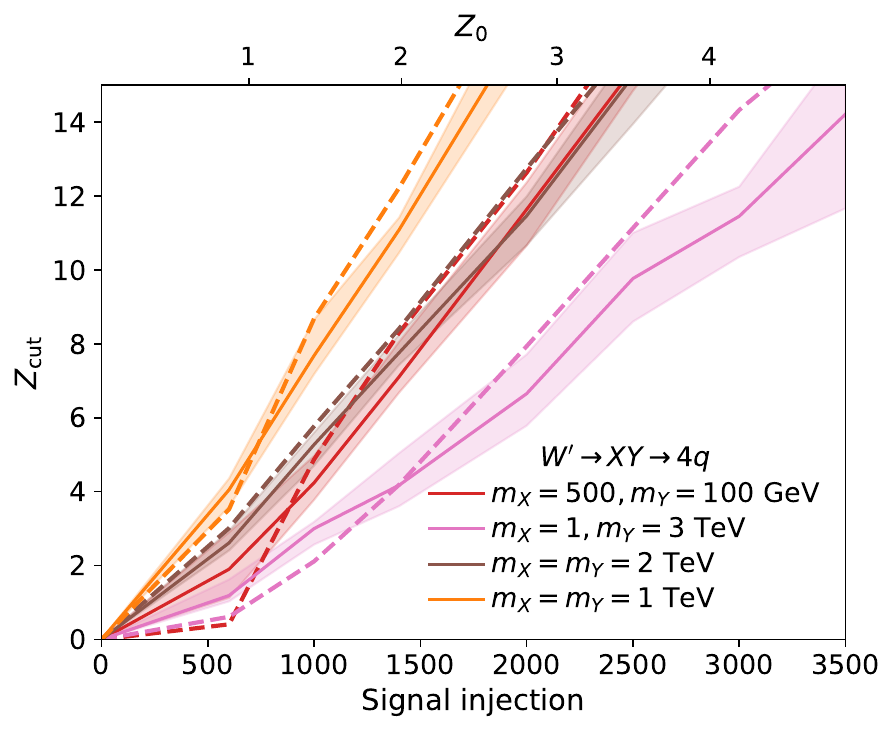}
    \caption{%
        Significance reached by IADs (dashed lines) and CATHODE (solid lines with 1$\sigma$ uncertainty bands) trained on all considered
        signal models as a function of the number of signal events. The $y$-axis shows the significance after the anomaly cut, \Z{\mrsd}{\mathrm{cut}}, while the top-$x$-axis indicates the initial significance \Z{\mrsd}{0}. All $G_\mathrm{KK}$ signals are shown on the left, $W'$ signals are shown on the right.
    }       
    \label{fig:bb3_single}
\end{figure}

The results are shown in Fig.~\ref{fig:bb3_single}.
We find that the IAD yields an improved significance for all considered signal models,
and is in most cases closely followed by CATHODE.
For the four-jets model (\wpxy with $m_X=m_Y=\SI{1}{\TeV}$), both the IAD and CATHODE
turn a $1\sigma$ excess into a discovery-level significance.
For the dijet $\gkk\to qq$ model, the IAD reaches $5\sigma$ using a bit more than
$2\sigma$ inclusive, while CATHODE yields no significant improvement. The IAD falling below CATHODE at low injections is due to statistical fluctuation, since the uncertainty band from the IAD is not shown here.

To look at the results from a different angle, we show in Table~\ref{tab:num-events-ad} the number of events needed to reach the 3 and $5\sigma$ when using the IAD or CATHODE, compared to the inclusive case without anomaly detection.
The IAD and CATHODE improve over the inclusive analysis for all signal models,
requiring between \num{25} and \SI{80}{\percent} fewer signal events to reach the
$5\sigma$ threshold.
This demonstrates the generality of feature sets based on event-level variables,
as also proposed in Ref.~\cite{Brennan:2025fqy}.
We note that our results only provide a lower bound and further improvements can
be achieved using better feature sets, generative models, or classifiers.

We find that in general, the classifiers perform better for models with higher jet 
multiplicity.
The two dijet signals with asymmetric jet masses, 
$G_\mathrm{KK}\to gR\to 3g$ with light $R$ and \wpxy with relatively light $X$ and $Y$,
are the hardest to find.
The IAD performs significantly better for the $G_\mathrm{KK}\to gR\to gt\bar t$ signal at the
same $R$ mass, showing that jet multiplicity is not the only factor affecting the performance.
A similar comparison for the heavy-$R$ case shows little difference.

\begin{table}
    \centering
    \caption{%
        Number of signal events needed to reach the $3\sigma$ significance threshold for all signal models. In the case of \Z{inc}{0}, about \num{3000} signal events are required for the same. 
    }
    \label{tab:num-events-ad}
    \begin{tabular}{cm{200pt}@{\hspace{10pt}}c@{\hspace{10pt}}c@{\hspace{10pt}}c}
    \toprule
    \multicolumn{2}{l}{} & \multicolumn{3}{c}{\small Signal events required [$\times 10^3$]} \\
    \cmidrule(lr{1.5pt}){3-5}
    \#jets & Signal process & \Z{\mrsd}{0} & \Z{\mrsd}{\mathrm{IAD}} & \Z{\mrsd}{\mathrm{CATHODE}} \\

    \midrule
        2 & $G_\mathrm{KK}\to q\bar q$
            &  2.0 & 1.1 & 2.2\\
        2 & $G_\mathrm{KK}\to gR$, $R\to gg$ ($m_R=\SI{500}{\GeV}$)
            & 1.9 & 1.4 & 1.9 \\
        2 & $G_\mathrm{KK}\to gR$, $R\to t\bar t$ ($m_R=\SI{500}{\GeV}$)
            & 1.9  & 0.9 & 1.2 \\
        3 & $G_\mathrm{KK}\to gR$, $R\to gg$ ($m_R=\SI{2217}{\GeV}$)
            & 1.9 & 0.8 &  1.7\\
        3 & $G_\mathrm{KK}\to gR$, $R\to t\bar t$ ($m_R=\SI{2217}{\GeV}$)
            & 1.9 & 0.8 &  1.4\\
        2 & \makecell[l]{$W'\to XY\to 4q$ \\ ($m_X=\SI{500}{\GeV}$, $m_Y=\SI{100}{\GeV}$)}
            &  2.1 & 0.8 & 0.7\\
        4 & $W'\to XY\to 4q$ ($m_X=m_Y=\SI{1}{\TeV}$)
            & 1.9 & 0.5 &  0.4\\
        4 & $W'\to XY\to 4q$ ($m_X=m_Y=\SI{2}{\TeV}$)
            & 2.0 & 0.6 &  0.6\\
        4 & $W'\to XY\to 4q$ ($m_X=\SI{1}{\TeV}$,
                          $m_Y=\SI{3}{\TeV}$)
            & 2.0 & 1.1 &  1.0\\

        \bottomrule
    \end{tabular}
\end{table}

\subsection{Mixed Decays} 
\begin{figure}[h]
    \centering
    \includegraphics[width=0.5\textwidth]{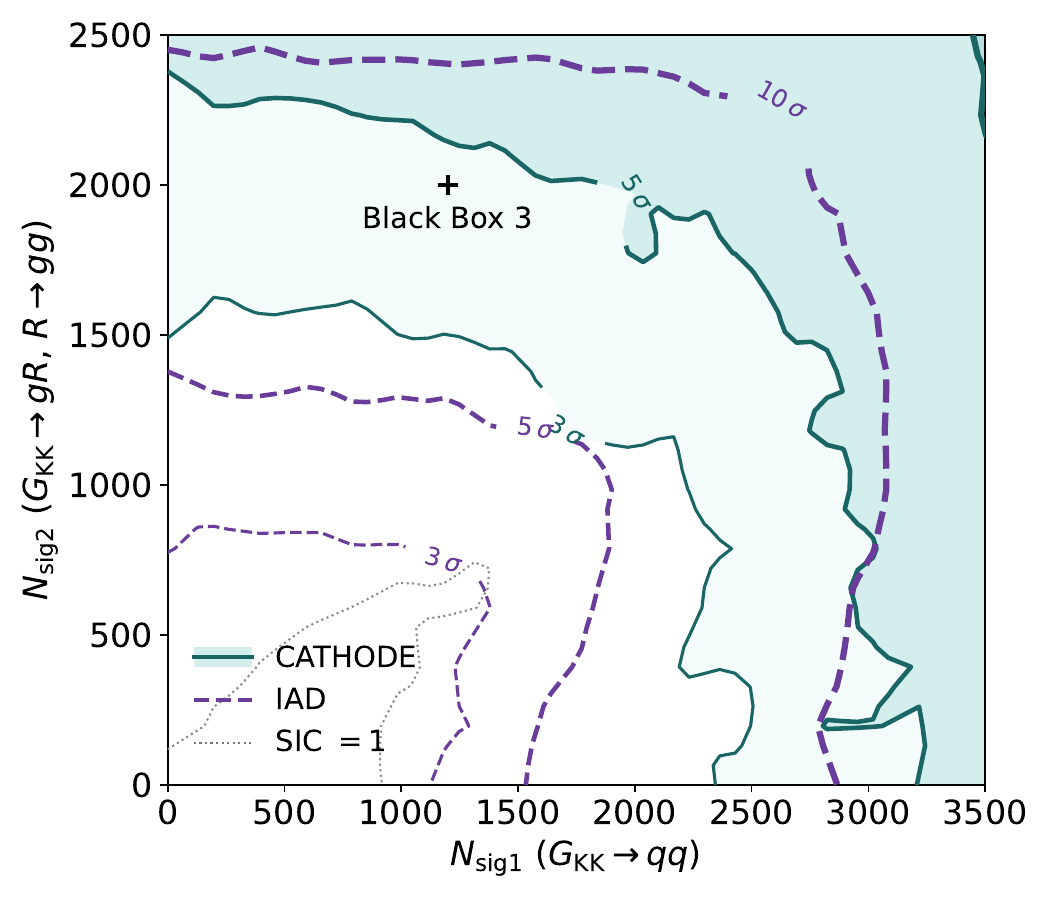}\relax
    \includegraphics[width=0.5\textwidth]{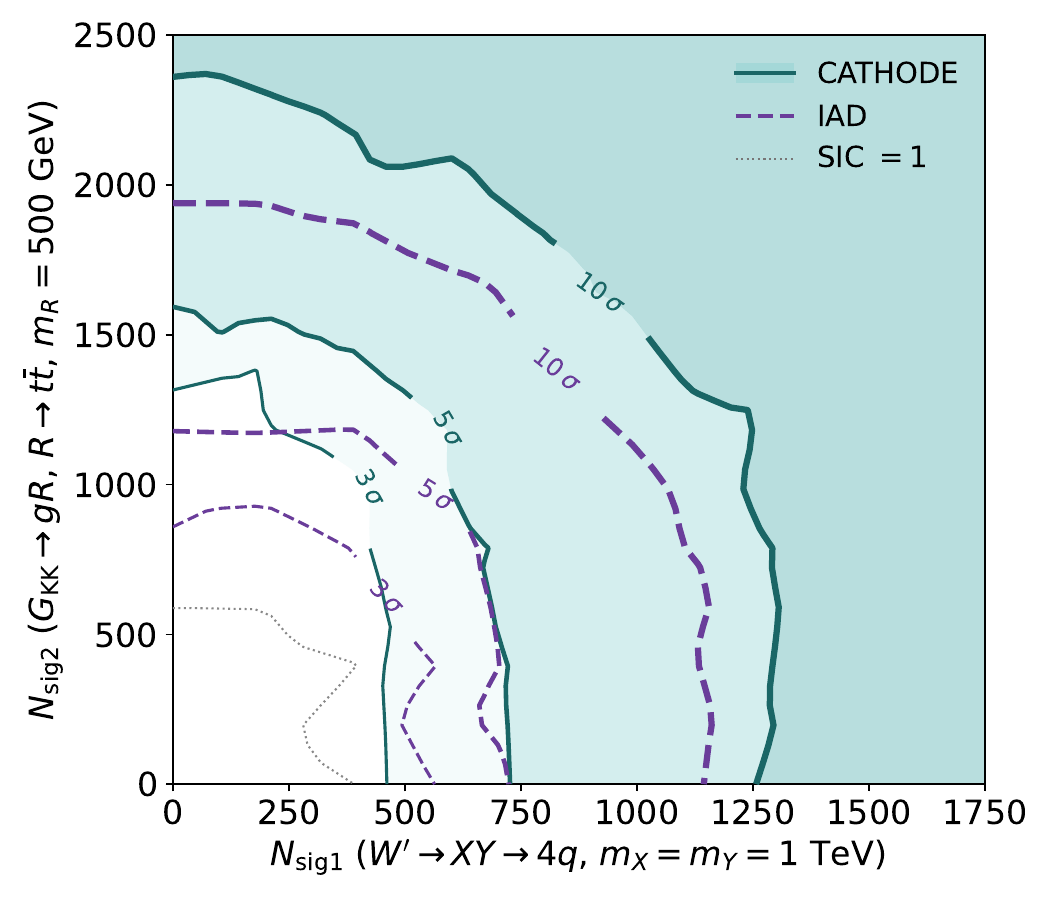}
    \caption{%
        Significance reached with IADs (dashed) and CATHODE classifiers (lined) for different mixing
        scenarios: the two signals from the Black Box 3 dataset (left), and the
        $W'\to XY\to 4q$ with $m_X=m_Y=\SI{1}{\TeV}$ and the
        $G_\mathrm{KK}\to gR\to gt\bar t$ with $m_R=\SI{500}{\GeV}$
        (right).
        The black cross shows the signal composition in the Black Box~3 dataset.
        The gray dotted line corresponds to a SIC of 1 with the IAD.
        The 3 (thin), 5 (medium), and $10\sigma$ (thick) contours are shown for both the IAD and CATHODE. 
    }
    \label{fig:weak-mix}
\end{figure}

We now turn to a signal with multiple decay modes, as in Black Box 3 dataset, which we construct by mixing several of our signal samples. We train these classifiers with a common label for both signals with mix of varying fractions.
For weakly supervised classifiers, the significance depends on the absolute number of events
present in the dataset. 
In Fig.~\ref{fig:weak-mix}, we present scans of the median significance obtained for the signal mixings with various fractions of both signals as well as the amount of signal needed to reach
the 3 and $5\sigma$ significance levels.
For both IADs and CATHODE, classifiers are individually trained and evaluated on the various signal fraction mixes. The beyond 3 and 5$\sigma$ regions for CATHODE are shown as bands. As seen in the single signal decay, a similar gap between IAD and CATHODE is observed.

We achieve a significance exceeding $5\sigma$ ($3\sigma$) with the IAD (CATHODE) for the signal in the original Black Box 3 dataset, which corresponds to a significance gain of a factor 1.6 with an IAD and 0.8 with the CATHODE method relative to \Z{\mrsd}{0}, for our signal region choice.
Compared to the inclusive significance \Z{inc}{0}, this represents total
significance gains by factors of 2.4 and 1.2 for the IAD and CATHODE, respectively.

We also observe that in the presence of two overlapping anomalies, such as for
a particle with multiple decay modes, the total significance achieved by CATHODE
or the IAD is not a simple combination of the significances of the two signals
taken independently.
Indeed, as seen in Fig.~\ref{fig:weak-mix}, the total significance is better
captured by the maximum of the individual significances rather than their sum.
This is observed both with Black Box 3 and $W'\to XY\to 4q$ signals.
We leave the understanding of this effect to future work.

\subsection{R\&D Dataset}

In this section, we cross-check the applicability of our method on the RnD dataset.
We evaluate our new mass definition on the R\&D dataset along with the feature set previously tested in the CATHODE implementation, optimized for dijet signal resonant at $\mjj=\SI{3.5}{\TeV}$. We utilize it as a benchmark to compare the two mass definitions and the feature sets. We run CATHODE and IADs for the signal regions  $3.3<\mjj<\SI{3.7}{\TeV}$ and $2.7<\mrsd<\SI{3.7}{\TeV}$, which both contain about 80\% of the signal events. We again evaluate the performance of CATHODE and IAD for various signal injections and show the median significance achieved by training \SI{10} classifiers at the background efficiency of \SI{1}\percent.

\begin{figure}[h]
    \includegraphics[width=0.5\textwidth]{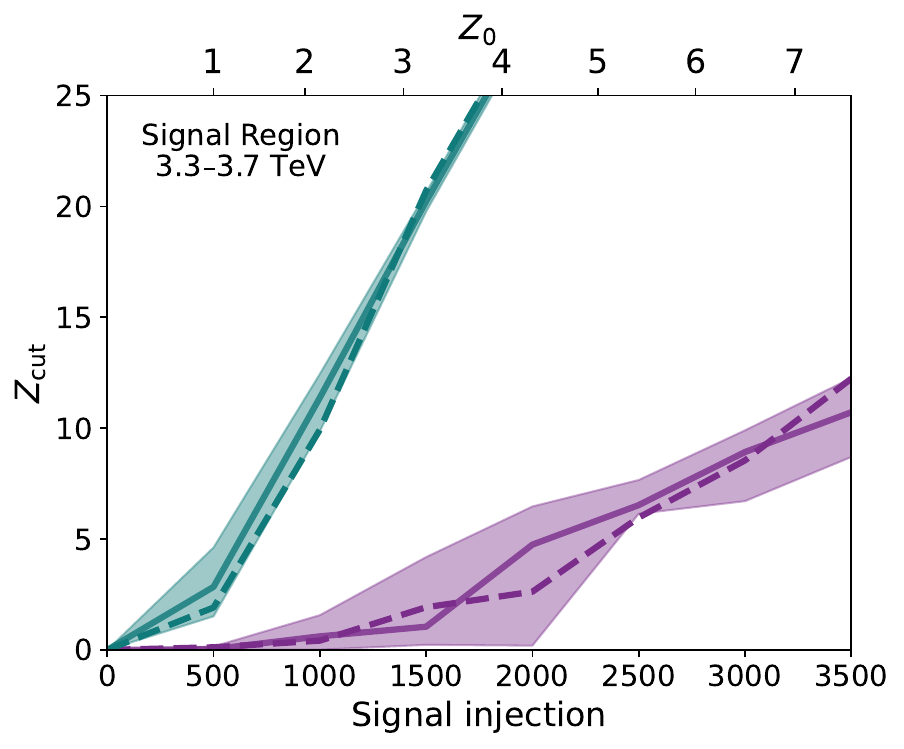}\relax
    \includegraphics[width=0.5\textwidth]{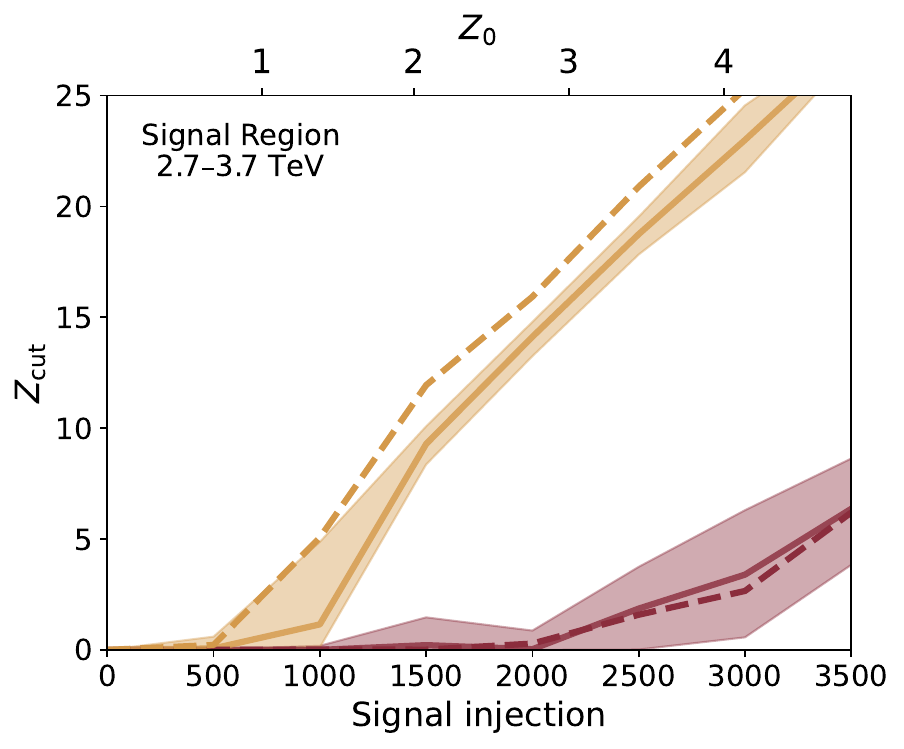}\par
    \vspace{4mm}
    \centering
    \includegraphics[width=0.7\textwidth]{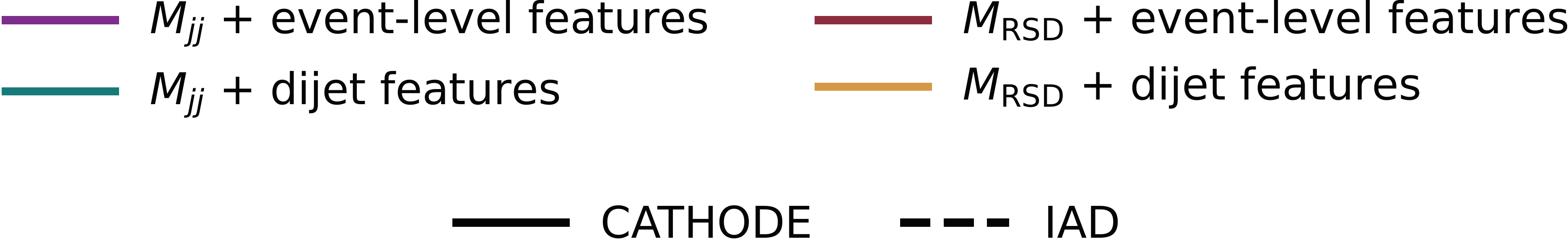}
    \vspace{2mm}
    \caption{%
        Significance reached with the IAD (dashed lines) and CATHODE (solid lines with 1$\sigma$ uncertainty bands) trained on \mjj (left) and \mrsd (right) with both feature sets for the dijet signal of the R\&D dataset resonant at $\SI{3.5}{\TeV}$.        
}
    \label{fig:rnd}
\end{figure}

The results are shown in Fig.~\ref{fig:rnd}. We show \mjj (left) and \mrsd (right) as the resonant feature paired with both feature sets. We observe that the dijet feature set gives us better performance in both cases as expected in comparison to the event-level feature set. For \mjj and the dijet feature set, although this set is the same as the previous implementation~\cite{Hallin:2021wme}, the event processing and training procedure differ substantially, which accounts for the difference in performance. 
We observe that with this signal, the event-level feature set requires more signal events to achieve the same significance, for both mass definitions.

On the right, \mrsd achieves strong performance with dijet features despite having a wider signal region. It demonstrates that \mrsd paired with a sufficiently powerful feature set can serve as a good working point. A more optimal feature set can always be optimized in accordance with the specific research aims.

\section{Summary}
\label{sec:summary}

Anomaly detection is of growing importance in the quest for the discovery of new physics beyond the Standard Model.
In this work, we have investigated a strategy to further extend the sensitivity of resonant bump-hunts to scenarios with an arbitrary number of jets in the final state.

The first and straightforward observation is that while a dijet mass trivially fails for final states with three or four jets, a combined mass of all particles remains resonant, but is also too broad to define sensible signal regions for anomaly detection. Applying a classical method from jet substructure --- recursive soft drop --- to event-level mass maintains, after some tuning, this attractive property and produces at the same time sufficiently narrow peaks to enable resonant anomaly detection.
By itself, the choice of resonant mass increases the expected significance obtainable via a counting experiment by 50 to 80\% depending on the signal model as compared to the initial significance.

Next, we combine this improved resonant feature with weakly-supervised anomaly detection. A simple Idealized Anomaly Detector (IAD) approach using high-level features and assuming perfect background estimation in the signal region further reduces the amount of events required to observe a new signal. In the case of a four-jet signal, the number of signal events required to reach $3\sigma$ is reduced from \num{2000} to \num{800} events.

This gain is however diminished when carrying out the full CATHODE procedure, i.e., when training and evaluating a generative model interpolated from the side-band into the signal region. For a single decay mode, CATHODE does follow the IAD closely and we observe overall good performance across all signals.
As this gap just points out a deficiency in the generative model that can be remedied with further model hyperparameter tuning (or the exploration of alternative generative architectures) we leave the resolution of this aspect to a full experimental implementation of the proposed method.

Inspired by Black Box 3 of the LHC Olympics, we also consider scenarios where a resonance with different decay modes is present in the data. 
Interestingly, we observe that the resulting significance is not  affected by the present anomalous decay modes equally. 
Instead, the maximum significance remains essentially unchanged when a second signal component is added with improvements only occurring if that other signal component becomes dominant.
These results indicate that weakly supervised anomaly detection does not yet constructively combine multiple signatures, motivating further studies of e.g.\ clustered approaches.

However, utilizing the proposed combination of recursive soft-drop mass with anomaly detection, we achieve a significance exceeding 5$\sigma$ with the IAD for the signal in the original Black Box 3 dataset. This corresponds to a total significance gain of a factor 1.6 relative to \Z{\mrsd}{0} (for our signal region choice) and a factor of 2.4 relative to \Z{inc}{0}, respectively. The later 
can be compared to the significance improvement of 1.4 reported by Ref.~\cite{Matos:2024ggs} with the ANTELOPE method, the previous state of the art on Black Box 3, marking a gain of around seventy percent.

Overall, these results further emphasize the adaptability of resonant anomaly detection as a main search strategy for new physics at the LHC, especially as these results are expected to naturally combine with better feature representations and anomaly scores. A further interesting development, however beyond the scope of this paper, would be to obtain definitions of a resonant feature that also includes lepton or missing energy information.

\newpage
\section*{Acknowledgments}

This work was started by LM, SH, and GK at the PhysTeV workshop 2023 in Les Houches and we thank the organizers for that opportunity.
We especially acknowledge and thank Luigi Favaro for his contributions in the early phase of this project. We also thank Marie Hein, Lukas Lang, Michael Krämer, and Ranit Das for valuable discussions on sample generation, code, and anomaly detection in general.

We thank the authors of the \textsc{FastJet}~\cite{Cacciari:2011ma} library for their
excellent implementation of the jet algorithms used in this paper. 

\paragraph{Funding information}
This research was supported by the Maxwell computational resources operated at Deutsches Elektronen-Synchrotron DESY, Hamburg, Germany.
LM, CY, GK acknowledge support by the Deutsche Forschungs-
Gemeinschaft under Germany’s Excellence Strategy 390833306 – EXC 2121: Quantum Universe.  
The work of SHL and DS was also supported by the DOE under Award Number DOE-SC0010008.
The work of SHL was also supported by IBS under the project codes, IBS-R018\_ys-2026-a00 and IBS-R018-D1.
This research was supported in part by grant NSF PHY-2309135 to the Kavli Institute for Theoretical Physics (KITP).

\section*{Code}
The code for this paper can be found here ~\cite{multijet_anomaly_code}.

\begin{appendix}
\section{Statistical Model}
\label{sec:combine}

To estimate the significance obtained with the various mass definitions, we conduct fits based
on histograms of each mass distribution. In this Appendix, we describe the statistical 
modeling used in the fits, which relies on the \textsc{Combine} tool~\cite{CMS:2024onh}
version 10.3.3.

We start by constructing histograms $\{b_i\}$ and $\{s_i\}$ of the background and signal
mass distributions, with bins $i=1, 2, \dots, 500$. We adjust the range based on the
background distribution for each variable: we use
$\num{2.5}<\mjj<\SI{7}{\TeV}$,
$\num{3}<\mall<\SI{9}{\TeV}$, and
$\num{2}<\mrsd<\SI{7}{\TeV}$.
We verified that our results are stable under variations of the number of bins or the fitting
range.

We normalize the histograms to the number of events present in the BB3 dataset, \num{996800}
for background and \num{3200} for signal. We consider each bin as an independent counting
experiment with Poisson statistics, resulting in the following binned-likelihood function:
\begin{equation}
    \mathcal L(\vec x; \mu) = \prod_{i=1}^{500} \text{Poiss}(x_i; b_i + \mu s_i),
\end{equation}
where $x_i$ is the observation in bin $i$, $\mu$ is the signal strength, and
$\text{Poiss}(k; \lambda) = \lambda^k e^{-\lambda}/k!$ is the Poisson distribution for an
expected number of events $\lambda$.

Following LHC conventions, we construct the test statistic
$q_0(\vec x)=-2\log\mathcal L(\vec x; 0)/\mathcal L(\vec x; \hat\mu)$, where $\hat\mu$ is the
maximum likelihood estimate of the signal strength. In case $\hat\mu<0$, $q_0$ is set to 0.
The asymptotic distribution of $q_0$ is tractable analytically in the large-sample limit,
giving a simple expression for the significance, $Z = \sqrt{q_0}$~\cite{Cowan:2010js}.
We use this formula to obtain the median significance in the presence of signal ($\mu=1$),
which we report in Table~\ref{tab:significance-combine}.

\end{appendix}

\clearpage
\bibliographystyle{SciPost_bibstyle}
\bibliography{HEPML,references}

\nolinenumbers

\end{document}